\documentclass{article}

\usepackage{microtype}
\usepackage{graphicx}
\usepackage{booktabs} 

\usepackage{hyperref}

\usepackage[accepted]{icml2019}

\usepackage[utf8]{inputenc} 
\usepackage[T1]{fontenc}    
\usepackage{hyperref}       
\usepackage{url}            
\usepackage{booktabs}       
\usepackage{amsfonts}       
\usepackage{nicefrac}       
\usepackage{microtype}      

\usepackage{amsmath}
\usepackage{times}
\usepackage{epsfig}
\usepackage{tabularx}
\usepackage{graphicx}
\usepackage[labelfont=bf,textfont={bf}]{caption}
\usepackage[labelfont=bf,textfont={bf}]{subcaption}
\usepackage{color}
\usepackage{xcolor}
\usepackage{colortbl}
\usepackage{xspace}
\usepackage{thumbpdf}
\usepackage{listings}
\usepackage{verbatim}
\usepackage{authblk}

\usepackage[inline]{enumitem}

\newcommand{\name}{Aurora\xspace}

\newcommand{\sub}[1]{\vspace{-4pt}\noindent\textbf{#1 }} 

\makeatletter
\newcommand{\printfnsymbol}[1]{%
  \textsuperscript{\@fnsymbol{#1}}%
}
\makeatother

\begin{document}

\twocolumn[
\icmltitle{A Deep Reinforcement Learning Perspective on Internet Congestion Control}



\icmlsetsymbol{equal}{*}

\begin{icmlauthorlist}
\icmlauthor{Nathan Jay}{equal,uiuc}
\icmlauthor{Noga H. Rotman}{equal,huji}
\icmlauthor{P. Brighten Godfrey}{uiuc}
\icmlauthor{Michael Schapira}{huji}
\icmlauthor{Aviv Tamar}{technion}
\end{icmlauthorlist}

\icmlaffiliation{uiuc}{University of Illinois at Urbana-Champaign}
\icmlaffiliation{huji}{Hebrew University of Jerusalem}
\icmlaffiliation{technion}{Technion}

\icmlcorrespondingauthor{Nathan Jay}{njay2@illinois.edu}
\icmlcorrespondingauthor{Noga H. Rotman}{nogar02@cs.huji.ac.il}
\icmlkeywords{Machine Learning, Reinforcement Learning, Congestion Control, ICML}

\vskip 0.3in
]


\printAffiliationsAndNotice{\icmlEqualContribution} 



\begin{abstract}
We present and investigate a novel and timely application domain for deep reinforcement learning (RL): Internet congestion control. Congestion control is the core networking task of modulating traffic sources' data-transmission rates to efficiently utilize network capacity, and is the subject of extensive attention in light of the advent of Internet services such as live video, virtual reality, Internet-of-Things, and more. We show that casting congestion control as RL enables training deep network policies that capture intricate patterns in data traffic and network conditions, and leverage this to outperform the state-of-the-art. We also highlight significant challenges facing real-world adoption of RL-based congestion control, including fairness, safety, and generalization, which are not trivial to address within conventional RL formalism. To facilitate further research and reproducibility of our results, we present a test suite for RL-guided congestion control based on the OpenAI Gym interface.  

\end{abstract}

\section{Introduction}
\label{sec:intro}

Deep RL has gained substantial popularity in light of its applicability to real-world decision making, generating headlines by facilitating complex tasks like protein folding~\cite{alphafold} and beating human experts in challenging contexts such as Go~\cite{silver2016mastering}. 
We wish to capitalize on these advancements to tackle the crucial and timely challenge of Internet congestion control.

\subsection{Internet Congestion Control}

In today's Internet, multiple network users contend over scarce communication resources. Consequently, the data-transmission rates of different traffic sources must be modulated dynamically to efficiently utilize network resources and provide good user experience. This challenge is termed ``congestion control'' and is fundamental to computer networking research and practice. Indeed, congestion control has a crucial impact on user experience for Internet services such as video streaming and voice-over-IP, as well as emerging Internet services such as augmented and virtual reality, Internet of Things, edge computing, and more.

\begin{figure}[t]
\centering
\includegraphics[width=\columnwidth, trim={0 5cm 0 5cm},clip]{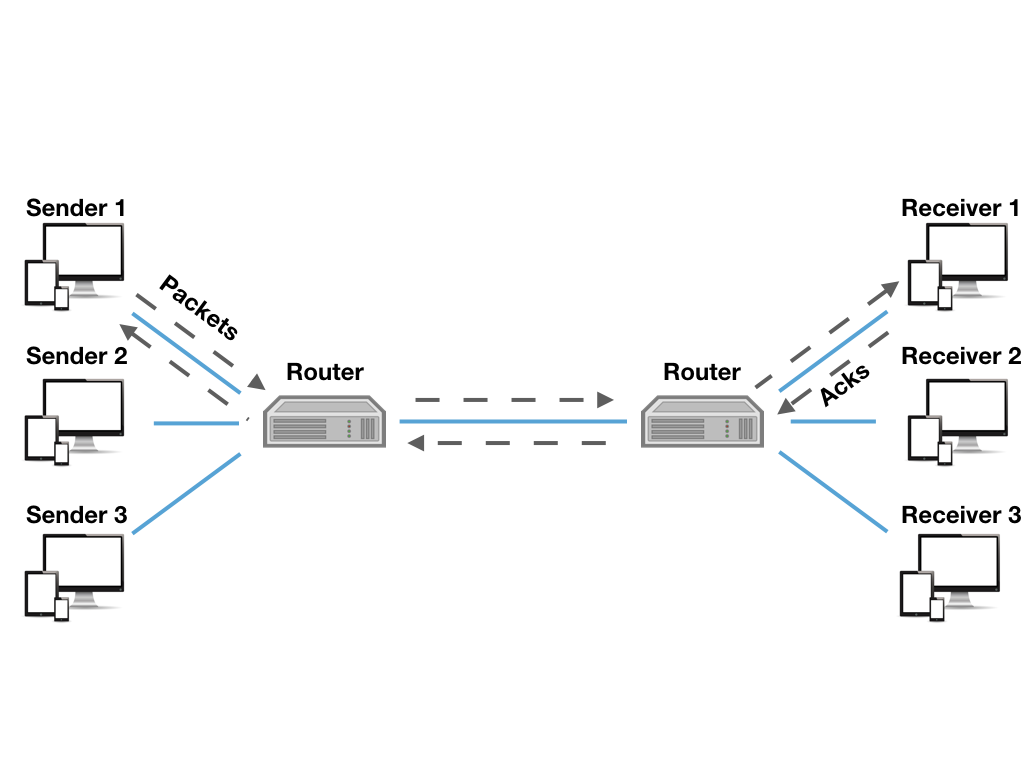}
\caption{Multiple traffic flows sharing a link}
\label{fig:link}
\end{figure}

Consider the illustrative example in Figure~\ref{fig:link}, which depicts multiple \emph{connections} (also referred to as ``\emph{flows}'') sharing a \emph{single} communication link. Each connection consists of a traffic sender and a traffic receiver. The sender streams data packets to the receiver and constantly experiences feedback from the receiver for sent packets in the form of packet acknowledgements (ACKs). The sender adjusts its transmission rate in response to this feedback. The manner in which the sending rate is adjusted is determined by the \emph{congestion control protocol} employed by the connection's two end-points. The interaction of different connections gives rise to network dynamics, derived from the connections' congestion control protocols, the link's capacity (bandwidth), and also the link's buffer size and \emph{packet-queueing policy}, which determine how (and whose) excess traffic is discarded.

Even this simple single-link scenario illustrates the complexity of congestion control, in which different connections select rates in an uncoordinated, decentralized manner, typically with no explicit information about competing connections (number, times of entry/exit, employed congestion control protocols), or the network (link's bandwidth, buffer size, and packet-queuing policy). Naturally, these challenges are exacerbated when traffic is forwarded across multiple links, and since networks greatly vary in sizes, link capacities, network latency, level of competition between connections, and more. Consequently, even after three decades of research on Internet congestion control, heated debates linger about the ``right'' approaches \cite{schapira2017congestion}.

\subsection{Motivating RL-Guided Congestion Control}

We argue that investigating RL in the context of congestion control is important in two respects: (1) improving the performance of a crucial component of the Internet's communication infrastructure, with the potential of impacting the user experience of essentially every performance-sensitive Internet service, and (2) providing an exciting new ``playground'' for RL schemes that poses novel real-world-motivated research challenges.

A congestion control protocol can be regarded as mapping a locally-perceived history of feedback from the receiver, which reflects past traffic and network conditions, to the next choice of sending rate. We hypothesize that such local history contains information about patterns in traffic and network conditions that can be exploited for better rate selection by \emph{learning} the mapping from experience via a deep RL approach. This hypothesis is motivated by the recent success of deep learning in extracting useful features across various domains, including image, speech, and games.

To illustrate the types of patterns that can be learned by RL schemes, consider the following two examples:

\vspace{0.03in}{\bf Distinguishing non-congestion loss from congestion-induced loss.} A packet loss can be indicative of different phenomena~\cite{allegro,bbr,vivace}; some 
might be induced by congestion, i.e., result from exceeding the network capacity, whereas others 
might be non-congestion-related, e.g., due to handover between mobile base stations. A well-known deficiency of today's prevalent congestion control protocol, the Transmission Control Protocol (TCP), is its innate inability to distinguish between these two forms of packet loss, which can lead to highly suboptimal 
rate choices (\citealt{allegro}).

\begin{figure}[t]
    \centering
    \includegraphics[width=0.9\columnwidth,trim={0 0 0 1.45cm},clip]{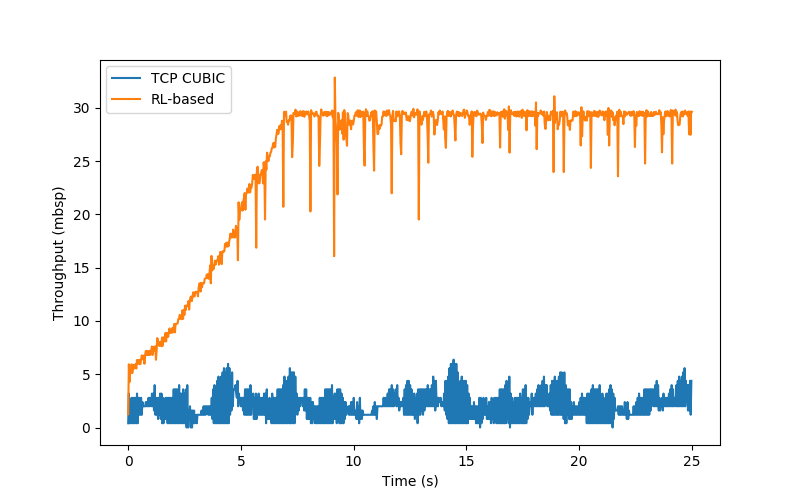}
    \caption{A 25-second trace of throughput for TCP-CUBIC and an RL-based protocol with 1\% random loss on a 30 Mbps bandwidth link and a 10-packet queue.}
    \label{fig:intro_random_loss}
\end{figure}

Figure \ref{fig:intro_random_loss} plots the 
throughput of a single flow on a link of bandwidth 30 Mbps, where $1\%$ of sent packets are randomly dropped, for two congestion control protocols: the CUBIC~\cite{ha2008cubic} variant of TCP, which is Linux's default; and a protocol designed within our RL framework.

Observe that TCP CUBIC, which halves the sending rate upon \emph{any} occurrence of loss, fails to fully utilize the link's bandwidth. In contrast, the RL-based protocol effectively distinguishes between the two types of losses; it often increases its rate upon encountering randomly-generated packet losses but avoids overshooting the link's capacity by ``too much'' by not increasing its rate when losses are due to congesting the link. Thus, the RL-based protocol is capable of maintaining high link utilization. Intuitively, distinguishing between these two types of losses is possible since random loss is unaffected by the sender's actions, whereas congestion-induced loss rate increases with the sending rate. 

\vspace{0.03in}{\bf Adapting to variable network conditions.} Network conditions, such as available link capacities, packet loss rates, and end-to-end latency, might be highly dynamic and change considerably over time (e.g., in mobile/cellular networks, \citealt{sprout}). TCP is notoriously bad at adapting to variable network conditions~\cite{allegro,bbr,vivace, copa}.

\begin{figure}[h]
    \centering
    \includegraphics[width=0.9\columnwidth,trim={0 0 0 1.45cm},clip ]{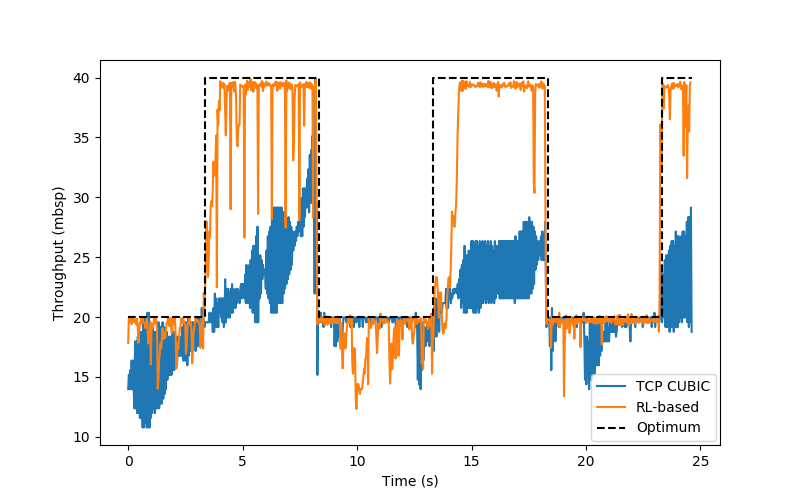}
    \caption{A 25-second trace of the throughput of TCP-CUBIC and an RL-based protocol on a link which alternates between 20 and 40 Mbps every five seconds.}
    \label{fig:intro_adaptability}
\end{figure}

Consider a single flow on a link whose capacity alternates between 20 Mbps and 40 Mbps every 5 seconds (with no random loss). 
Ideally, the congestion control protocol would change its sending rate so as to closely match the link's bandwidth, as illustrated by the dashed black line in Figure~\ref{fig:intro_adaptability}, thus utilizing the capacity without causing congestion. As shown in the figure, TCP CUBIC fails to do so. In contrast, a protocol designed within our RL framework closely approximates the desired behavior. Intuitively, this protocol learns, e.g., that sudden steep rises in packet loss indicate that available bandwidth has decreased considerably, and that no packet loss when the sending rate exceeds 20 Mbps indicates that the available bandwidth is higher (40 Mbps).

\subsection{Our Contribution}

We formulate a novel framework for RL-based congestion control protocol design, which extends the recently introduced Performance-oriented Congestion Control (PCC) approach~\cite{allegro,vivace}. We discuss challenges involved in casting congestion control as an RL task. We also describe remaining challenges facing the real-world adoption of RL-based congestion control schemes, such as fairness, safety, and generalization, which are not trivial to address within conventional RL formalism.

We utilize our framework to design \name. \name employs deep RL~\cite{sutton1998reinforcement,trpo} to generate a policy for mapping observed network statistics (e.g., latency, throughput) to choices of rates. Our preliminary evaluation results suggest training \name in simple, simulated environments is sufficient to generate congestion control policies that perform well also in very different network domains and which are comparable to, or outperform, recent state-of-the-art handcrafted protocols.


Our code\footnote{\href{https://github.com/PCCproject/PCC-RL}{https://github.com/PCCproject/PCC-RL}} is open-sourced as an OpenAI Gym environment and an accompanying testing module, to be used by RL researchers and practitioners to evaluate their algorithms. 



\section{RL Approach to Internet CC}
\label{sec:rl_approach}

We next provide a high-level overview of RL and  explain how congestion control can be formulated as an RL task.

\subsection{Background: Reinforcement Learning}
\label{subsec:background}

In RL~\cite{sutton1998reinforcement}, an \textit{agent} solves a sequential decision making problem by interacting with an \textit{environment}. 


At each discrete time step $t\in{0,1,...}$, the agent observes a (locally perceptible) state of the environment $s_t$, and selects an action $a_t$.
At the following time step $t+1$, the agent observes a \textit{reward} $r_t$, representing its loss/gain after time $t$, as well as the next state $s_{t+1}$.\footnote{In our setting, the next state is drawn from the environment's transition dynamics, taking into account the agent's action, the actions of other agents, and parameters unavailable to the agent, e.g. link capacity. This makes the problem an instance of a partially observable Markov decision process~\cite{kaelbling1998planning}.} 
The agent's goal is to choose a policy $\pi$ mapping states to actions that maximize the \textit{expected cumulative discounted return} $R_t=\mathbb{E}\big[\sum_{t}\gamma^{t}\cdot r_t\big]$, for $\gamma \in \big[0,1\big)$. The parameter $\gamma$ is termed the \textit{discount factor}. For large or continuous state and action spaces, this problem is intractable, and recent advances in deep RL employ deep neural networks to approximate the optimal $\pi$~\cite{trpo,silver2017mastering}.

\subsection{Congestion Control as RL}\label{subsec:cc_as_rl}

We formulate congestion control as a sequential decision making problem under the RL framework.

\sub{Actions are changes to sending rate.}
In our formulation, the agent is the sender of traffic and its actions translate to changes in sending rates. To formalize this, we adopt the notion of \emph{monitor intervals} (MIs) from~\cite{allegro,vivace}. Time is divided into consecutive intervals. In the beginning of each MI $t$, the sender can adjust its sending rate $x_t$, which then remains fixed throughout the MI. After experimenting with several options, we chose to express actions as \emph{changes} to the current rate (see Section~\ref{subsec:design_architecture}).\footnote{While our action formulation (periodic rate changes) is less nuanced than allowing the sender to choose the exact timing of each packet transmission, such a formulation is too expensive to realize with today's transmission speeds.}

\sub{States are bounded histories of network statistics.} After the sender selects rate $x_t$ at MI $t$, it observes the results of sending at that rate and computes a statistics vector $v_t$ 
from received packet-acknowledgements. 
We restrict our attention below to \emph{statistics vectors} consisting of the following:
\begin{enumerate*}[label=(\roman*)]
\item latency gradient~\cite{vivace}, the derivative of latency with respect to time;
\item latency ratio~\cite{remy}, the ratio of the current MI's mean latency to minimum observed mean latency of any MI in the connection's history; and
\item sending ratio, the ratio of packets sent to packets acknowledged by the receiver
\end{enumerate*}.

Networks greatly vary in terms of available bandwidth, latency, and loss rate. Our choice of elements comprising the statistics vector is intended to improve the generalization of our models by avoiding statistics that are expected to be highly variable across connections for no better reason than variation in link properties (e.g., the absolute value of experienced latency in milliseconds).



The agent's selection of the next rate change is a function of a \emph{fixed-length history} of the above statistics vectors collected from packet acknowledgements sent by the receiver. Considering a bounded-length history, instead of just the most recent statistics, allows our agent to detect trends and changes in network conditions and react more appropriately. Thus, the state at time $t$, $s_t$, is defined to be: $$s_t=(v_{t-(k + d)},\ldots,v_{t-d}),$$ for a predetermined constant $k>0$ and a small number $d$ representing the delay between choosing a sending rate and gathering results. We discuss how the length of the history, i.e., $k$, affects performance in \autoref{subsec:choices}.

\sub{Setting rewards.} The reward resulting from sending at a certain rate at a certain time may depend on the performance requirements of the specific application; some applications (e.g., online gaming) might require very low latency while for others (e.g., large file transfers) high bandwidth is much more crucial; some services might prefer low-but-constant bandwidth (no ``jitter''), while others may desire higher bandwidth and be more tolerant to bandwidth variation. We discuss specific reward functions in \autoref{subsec:design_architecture}. 

The effects of an action (change in rate) could potentially have non-immediate consequences, e.g., sending at too fast a pace could overload buffers and result in future packet losses and delays. In RL, long-horizon decision making is captured via the discount factor $\gamma$. We discuss the impact of $\gamma$ in our framework in \autoref{subsec:choices}.

\subsection{Other Considered Approaches}

We considered alternative formulations of congestion control as a learning task (most notably as a bandits problem) and alternative model architectures (including linear models) prior to settling on the ones presented here. 

Our results (see \autoref{fig:discount_factor}) show that to learn a reasonable policy, the discount factor $\gamma$ cannot be too low (e.g., $\gamma$ should be at least $0.5$), with high discount factors ($\gamma=0.99$) resulting in much faster learning. This is in agreement with the sequential nature of the task, in which rewards might be delayed due to effects such as limited buffer size on the link and increase in latency as a consequence of increase in link occupancy.

In addition, training linear models resulted in much worse performance---noticeably worse than even a single layer neural network. We also tried simple random search and hill-climbing for linear models, whic ~\cite{mania2018simple} recently showed performs well on continuous Mujoco tasks, but these were not competitive in our context.

In our experiments (\autoref{sec:evaluation}), a discount of 0.99 resulted in much faster learning (though eventually reaching a performance similar to 0.5 - see \autoref{fig:discount_factor}), showing that the stronger signal from delayed rewards is important.

\section{Introducing \name}\label{sec:design}
In this section we introduce \name: a specific implementation of RL for congestion control, based on the formulation above, that achieves state-of-the-art results. Our code is available at our 
\href{https://github.com/PCCproject/PCC-RL}{github repo}.

\subsection{Architecture}\label{subsec:design_architecture}
\sub{RL inputs and outputs.} We choose to map our agent's output 
based on the statistic vectors discussed in Section~\ref{sec:rl_approach} to a change in sending rate $x_{t-1}$ according to:
\[   
  x_{t} = 
\left\{
\begin{array}{ll}
      x_{t-1} * (1 + \alpha a_t) & a_t\geq 0 \\
      x_{t-1} / (1 - \alpha a_t) & a_t< 0 \\
\end{array}
\right. \]
where $\alpha$ is a scaling factor used 
to dampen oscillations (we use $\alpha = 0.025$).


\sub{Neural network.}
Neural network architectures vary significantly and research suggests new architectures at an incredible rate, so choosing \emph{the} optimal architecture is impractical.
We show, however, that even a simple architecture, i.e., a small \textit{fully connected} neural network, produces good results. We tested several options for the number of hidden layers and number of neurons per layer and chose an architecture with two hidden layers composed of $32\rightarrow 16$ neurons and tanh nonlinearity. After training three replicas of each considered architecture, this one produced the highest average training reward and exhibited high performance throughout our evaluation process (see \autoref{sec:evaluation}).


\sub{Reward function.}
We trained \name with a linear reward function that rewards throughput while penalizing loss and latency.  
State-of-the-art PCC-Vivace~\cite{vivace} and Copa~\cite{copa} try to optimize reward functions with different exponents and logarithms of these components, but have similar goals (high throughput, low latency, with PCC-Vivace penalizing loss as well). We choose the following linear function instead:
\[10 * throughput - 1000 * latency - 2000 * loss\]
where $throughput$ is measured in packets per second, $latency$ in seconds, and $loss$ is the proportion of all packets sent 
but not acknowledged. 
The scale of each factor was chosen to force models to balance throughput and latency for our chosen training parameters. In Section~\ref{sec:evaluation} we discuss the objective functions (or lack thereof) for other algorithms where we demonstrate \name's throughput-latency tradeoff.


\subsection{Training}\label{subsec:training}

We train our agent in an open-source gym environment described in detail in Section~\ref{sec:gym_env}. This environment simulates network links with a range of parameters. Our model was trained using the PPO algorithm~\cite{ppo}, as implemented in the stable-baselines python package (based on~\citealt{baselines}).

\subsection{Choice of Parameters}\label{subsec:choices}


While many parameters affect the quality of our final model, we next discuss two significant parameter choices: history length, and discount factor.

\begin{figure}[h]
    \centering
    \includegraphics[width=0.875\columnwidth, trim={0 0.25cm 0 1.45cm},clip]{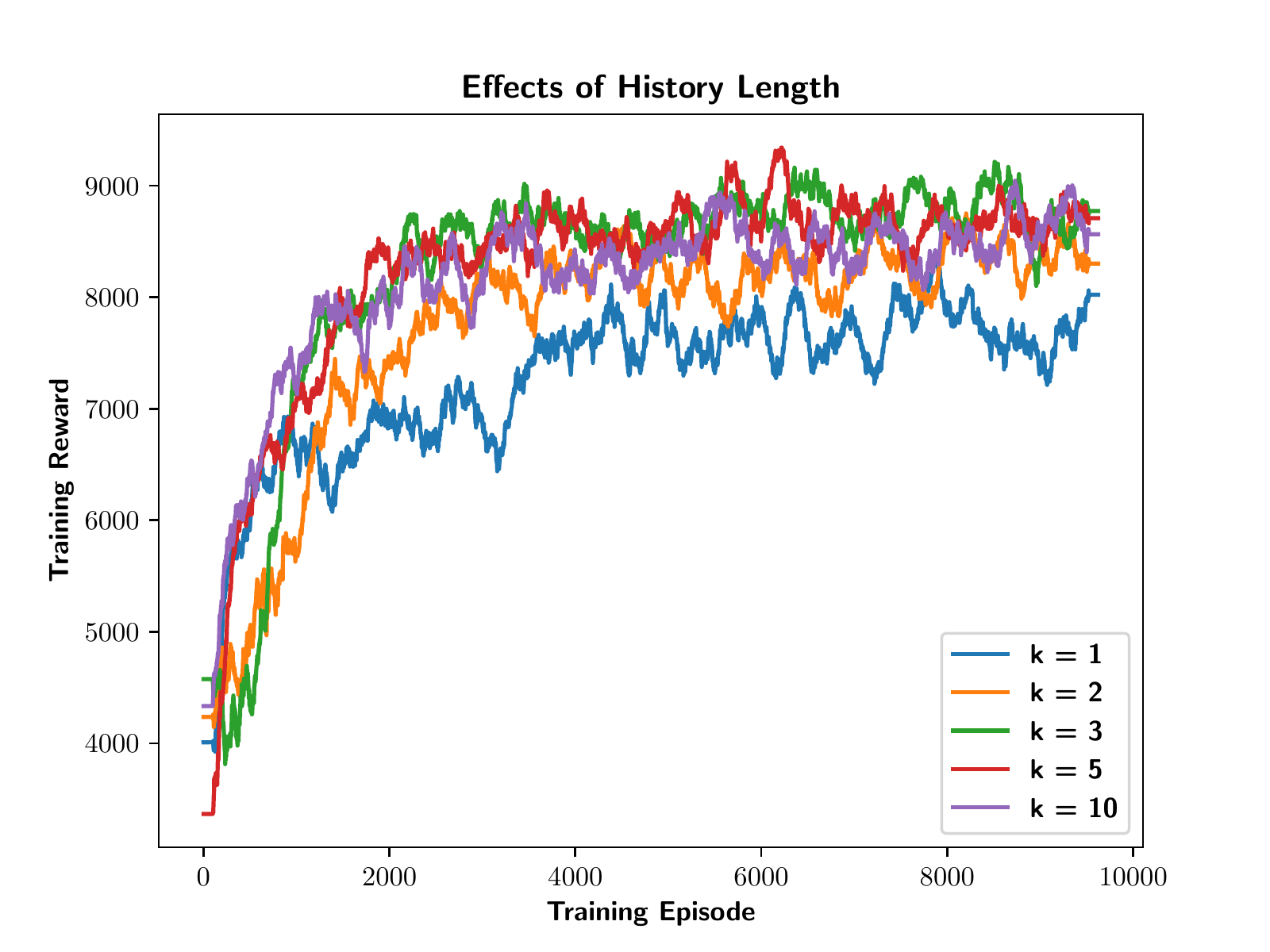}
    \caption{Training reward for agents with different values of history length. 
    Lines are smoothed averages of three models.
    }
    \label{fig:history_len}
\end{figure}

\sub{History length.}
A history length of $k$ means that the agent makes a decision based on the $k$ latest MIs worth of data. Intuitively, increasing history length should increase performance, as extra information is given. We trained models with $k$ ranging from 1 to 10 MIs. Figure~\ref{fig:history_len} shows the training reward of these models. Eventually, a model with $k=2$ was comparable to the model with $k=10$, but a model with a single history does not learn a comparable policy in our chosen number of training episodes.

\begin{figure}[!h]
    \centering
    \includegraphics[width=0.875\columnwidth, trim={0 0.25cm 0 1.45cm},clip]{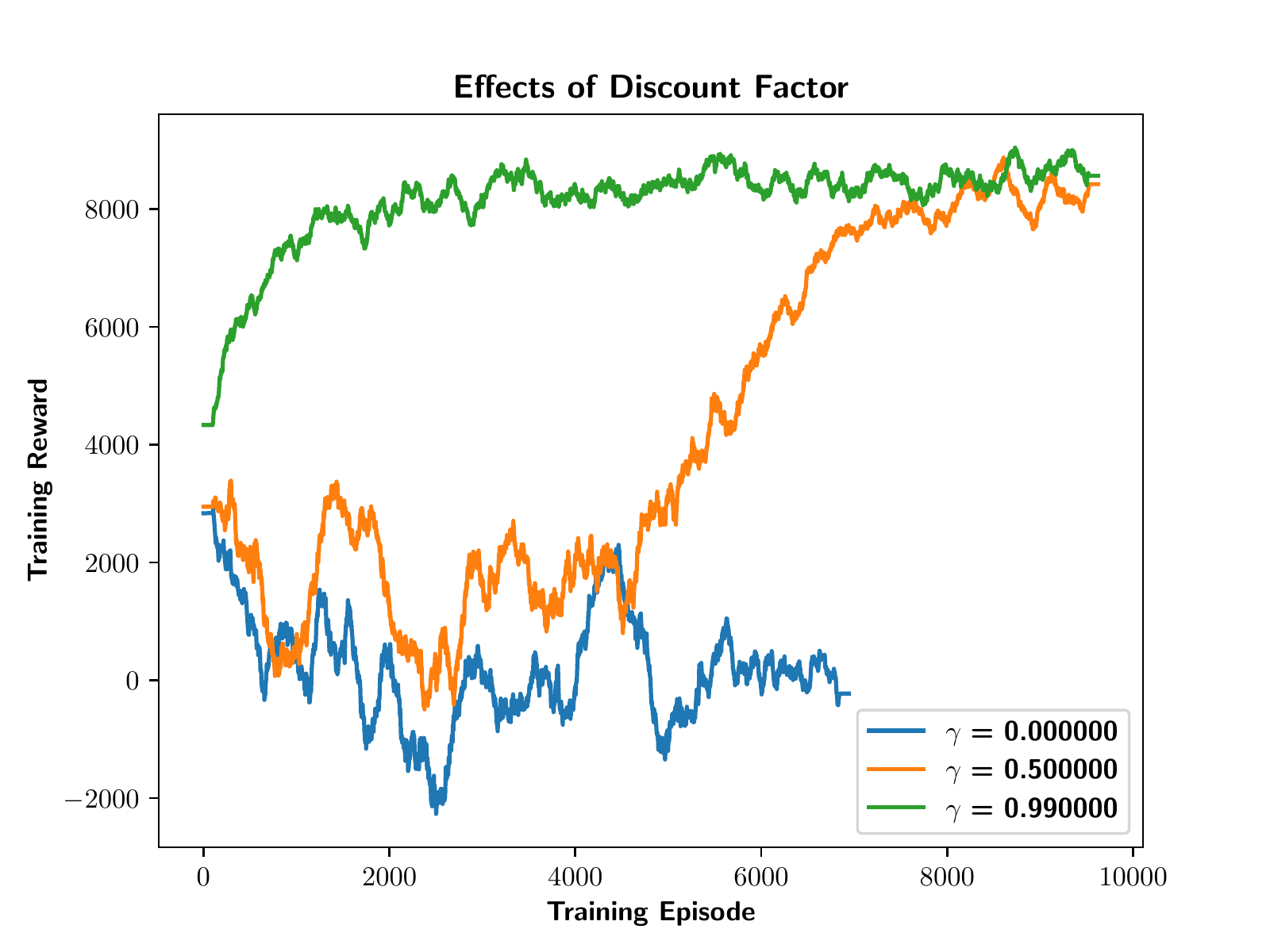}
    \caption{Training reward for agents with different values of gamma. Each line is the smoothed average of three models. The line for $\gamma = 0.00$ ends early because training exceeded a budgeted time.}
    \label{fig:discount_factor}
\end{figure}

\sub{Discount factor.}
We examined three different values of $\gamma$ and determined that $\gamma = 0.99$ gave the best results quickly, while $\gamma = 0.50$ eventually learned a reasonable policy, and $\gamma = 0.00$ failed to learn a useful policy. This seems intuitive, given the gap between taking an action and observing a reward. Figure~\ref{fig:discount_factor} shows these trends.

\section{Evaluation}\label{sec:evaluation}

Our training framework uses a simple simulation of a single traffic source on network links with varied parameters. Our testing suite, however, goes far beyond that---including link parameters outside the scope of the training values, dynamic links as opposed to the purely static links in training, and training in an emulated rather than simulated environment. In particular, we test the performance and robustness of our trained model using standard network research tools (Mininet~\cite{mininet} and Pantheon~\cite{pantheon}) for a wide variety of network scenarios, and provide comparisons to state-of-the-art congestion control protocols.

\subsection{Robustness}

\begin{figure*}[h]
\centering
    \begin{subfigure}{0.24\textwidth}
  \centering
  \includegraphics[width=\textwidth, trim={0 0 0 0},clip]{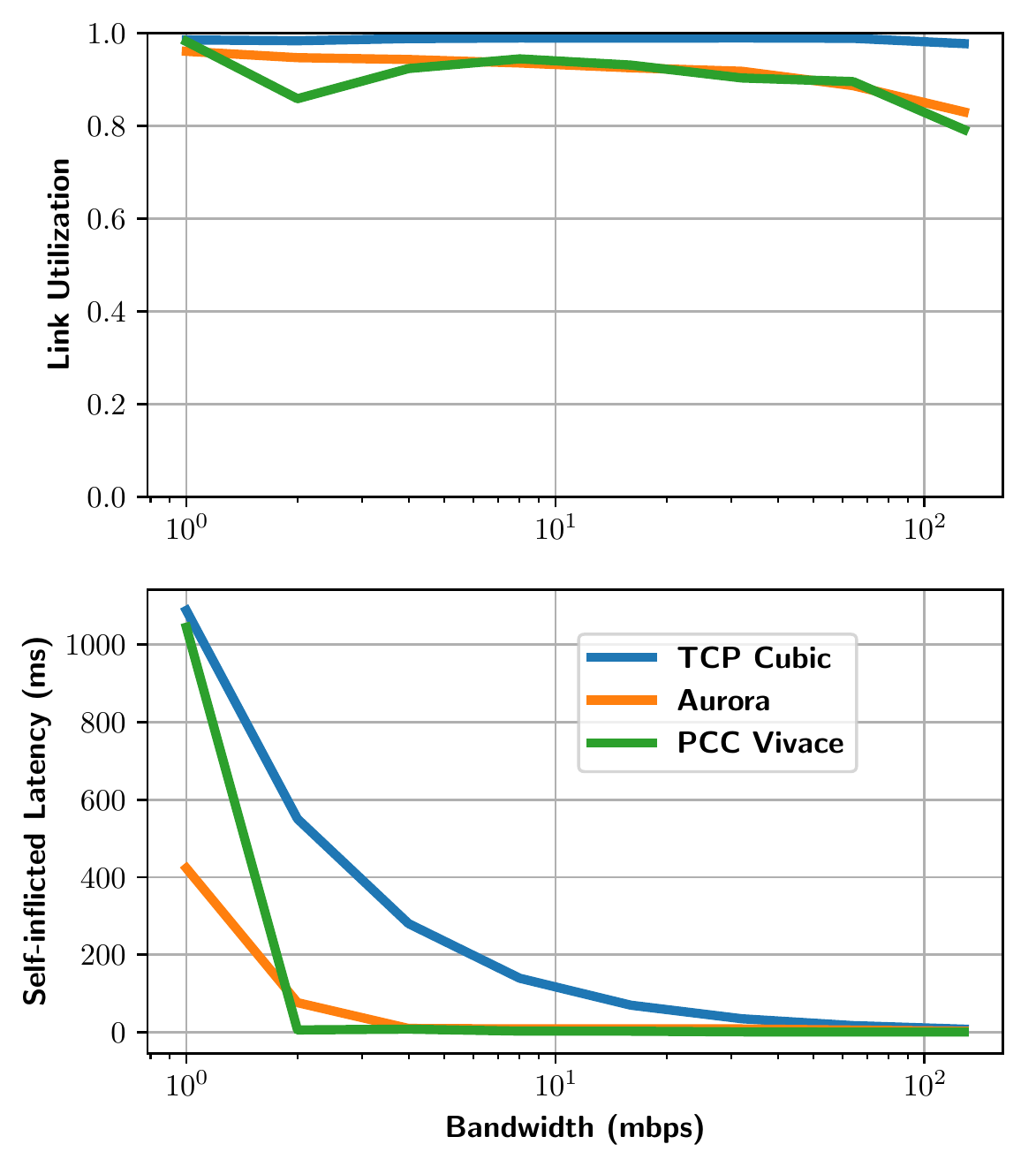}
    \caption{Bandwidth sensitivity}
    \label{fig:sensitivity_bandwidth}
\end{subfigure}    
\centering
\begin{subfigure}{0.24\textwidth}
  \centering
  \includegraphics[width=\textwidth, trim={0 0 0 0},clip]{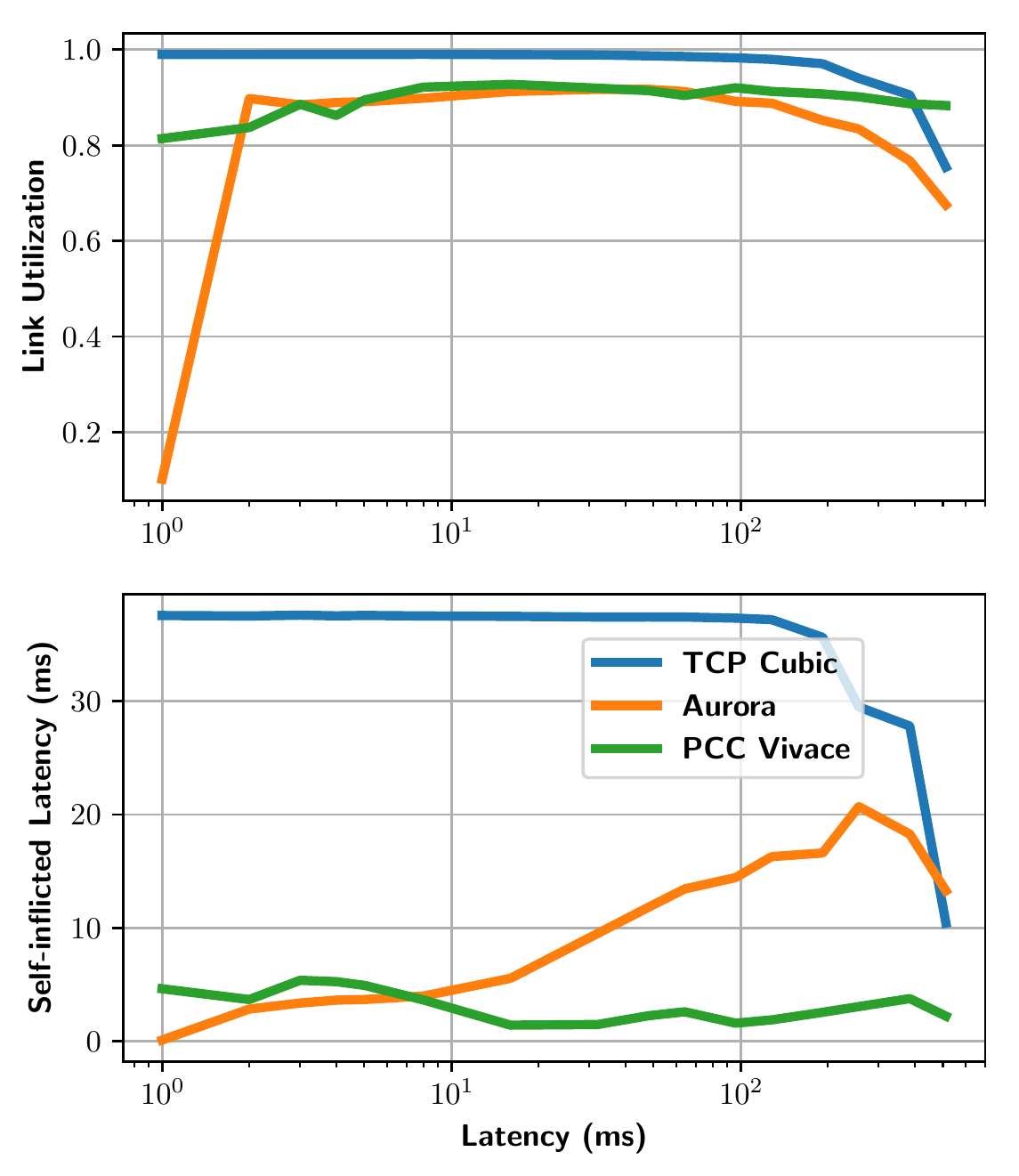}
    \caption{Latency sensitivity} \label{fig:sensitivity_latency}
\end{subfigure}
\centering
\begin{subfigure}{0.24\textwidth}
  \centering
  \includegraphics[width=\textwidth, trim={0 0 0 0},clip]{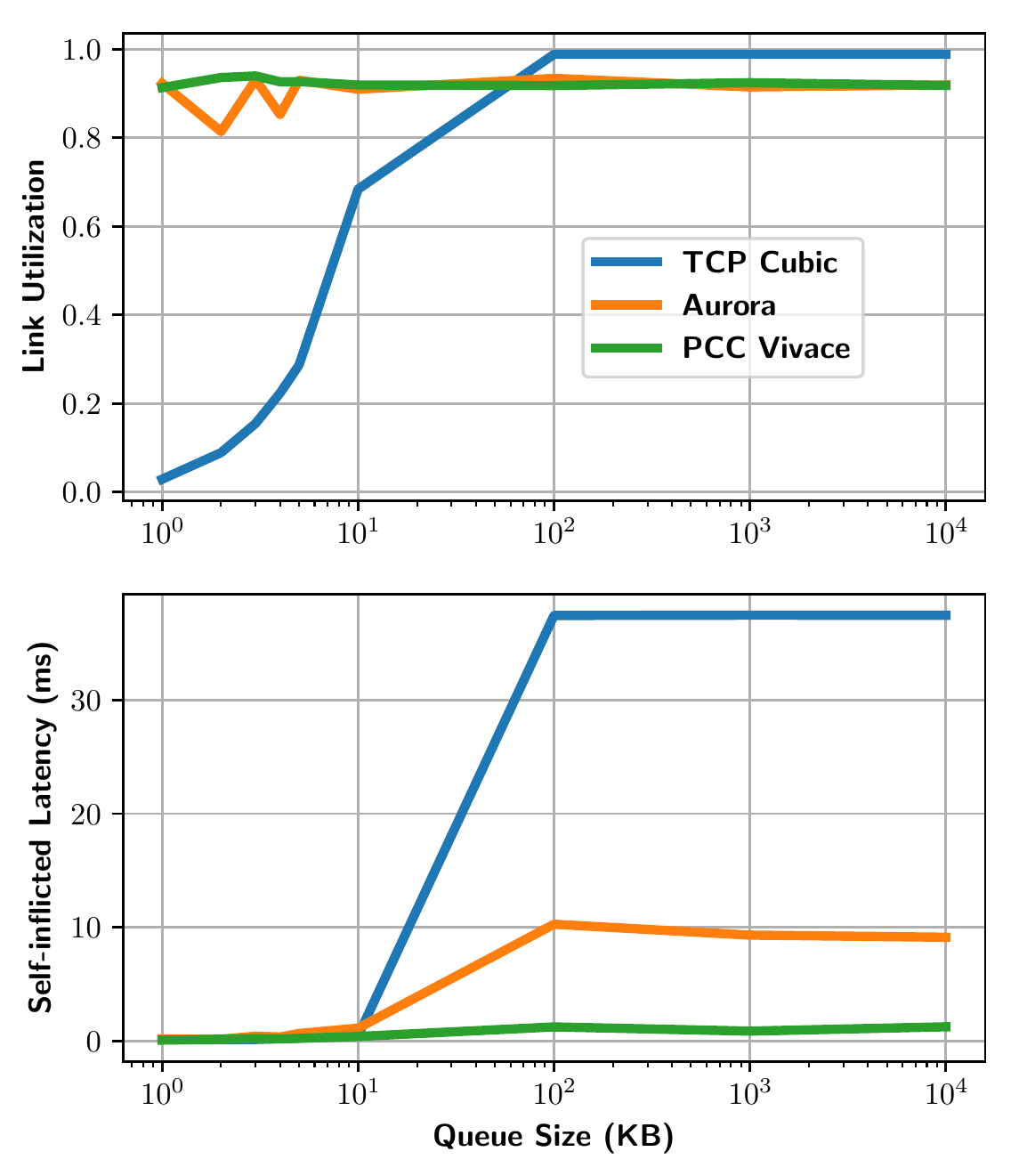}
    \caption{Buffer size sensitivity} \label{fig:sensitivity_buffer}
\end{subfigure}
\begin{subfigure}{0.24\textwidth}
  \centering
  \includegraphics[width=\textwidth, trim={0 0 0 0},clip]{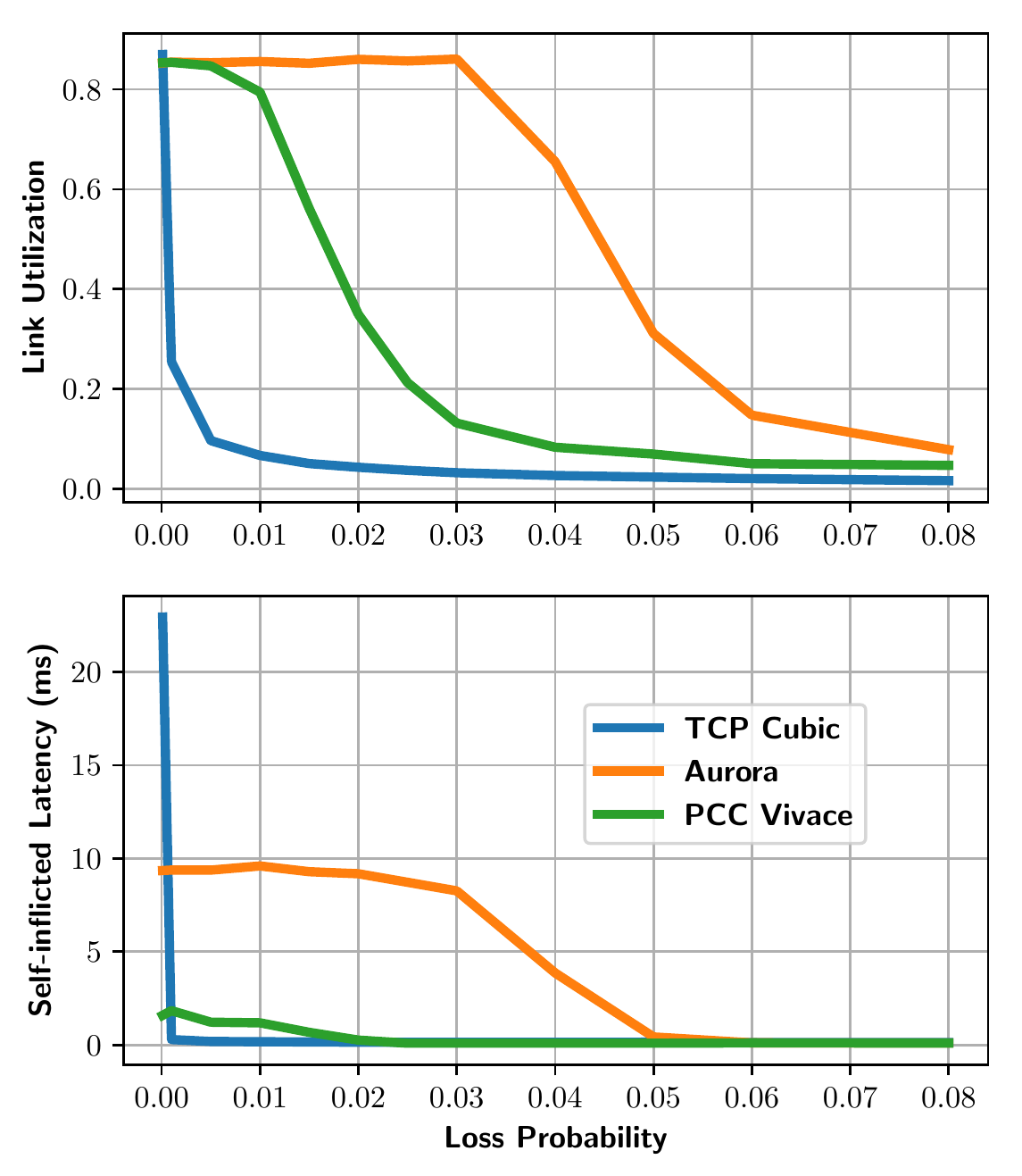}
    \caption{Loss rate sensitivity} \label{fig:sensitivity_loss_rate}
\end{subfigure}
\caption{Tests over a single link, showcasing the model's sensitivity to changes in bandwidth, latency queue size and loss rate. Throughput and self-inflicted latency are shown as the relevant parameter varies. In the top graph, higher is better. In the bottom graph, lower is better.}
\label{fig:sensitivity}
\end{figure*}

\autoref{fig:sensitivity} demonstrates our model's behavior when bandwidth, latency, queue size and random loss rate vary far outside of our training conditions. The Pantheon project
showed that varying these four link parameters is sufficient to emulate a wide variety of real-world network conditions. (Pantheon also includes a fifth parameter, setting queue behavior to fixed or exponentially-distributed service times, to better emulate mobile LTE links; this parameter is not available in Mininet.) We run each test for two minutes with a single sender over a single link. For each test, we compare the results of our model against TCP CUBIC (the current standard for congestion control), and PCC Vivace (\citealt{vivace}, a robust, state-of-the-art algorithm). Our supplemental material includes comparisons to several other state-of-the-art algorithms. Unless otherwise stated, Mininet links are configured with 30~Mbps capacity, 30~ms of latency, a 1000-packet queue, and 0\% random loss.

\sub{Bandwidth sensitivity.}
Our model was trained on links with bandwidth between 1.2~Mbps and 6~Mbps, but any reasonable congestion control should operate in a much wider range. In Figure~\ref{fig:sensitivity_bandwidth} we show that \name operates well between 1~Mbps and 128~Mbps, more than $20\times$ above its training environment. We start from our standard link configuration and configure the bandwidth differently for each test, ranging from 1 to 128~Mbps. Our model achieves near-perfect link utilization at all capacities, with latency far below TCP, and comparable to PCC-Vivace.

\sub{Latency sensitivity.}
We trained \name with link latency varying between 50ms and 500ms, but many Internet connections are expected to have latency much lower (and rarely higher) than this range. To test the robustness of our model on links with different latency, we vary the latency between 1~ms and 512~ms. \name performs poorly at 1ms latency because our emulation environment includes real packet processing, a process that adds noise on the order of around 1ms to the latency, but which was not present in training.

\sub{Queue sensitivity.}
\name was trained on links with queue size varying between 2 and 2981 packets. For this set of tests, we varied queue size between 1 and 10,000 packets. \autoref{fig:sensitivity_buffer} shows our model has throughput comparable to PCC-Vivace at all queue sizes and provided significant throughput improvements over TCP CUBIC in that range.

\sub{Loss sensitivity.}
We trained \name with random loss probability varying between 0\% and 5\%. This range already captures the full range of expected conditions for most Internet traffic, but we test up to 8\% random loss probability to explore the robustness of our model. In this test, \name provides near-capacity throughput at higher loss than either of the other algorithms.

\sub{Dynamic links.}
Some network links, particularly mobile links, have variable capacity that can depend on real-time changes in signal strength, whether from interference, distance or surroundings. The exact dynamics of link capacity vary widely, so we examine a simplified case. In this scenario, we consider a link whose capacity varies every five seconds to a newly chosen value distributed between 16~Mbps and 32~Mbps uniformly at random. We allow \name, along with five baseline schemes, to run at least five times on this dynamic link setup for two minutes per run. 

\begin{figure}[h]
    \centering
    \includegraphics[width=0.9\columnwidth,trim={0 0 0 1.45cm},clip]{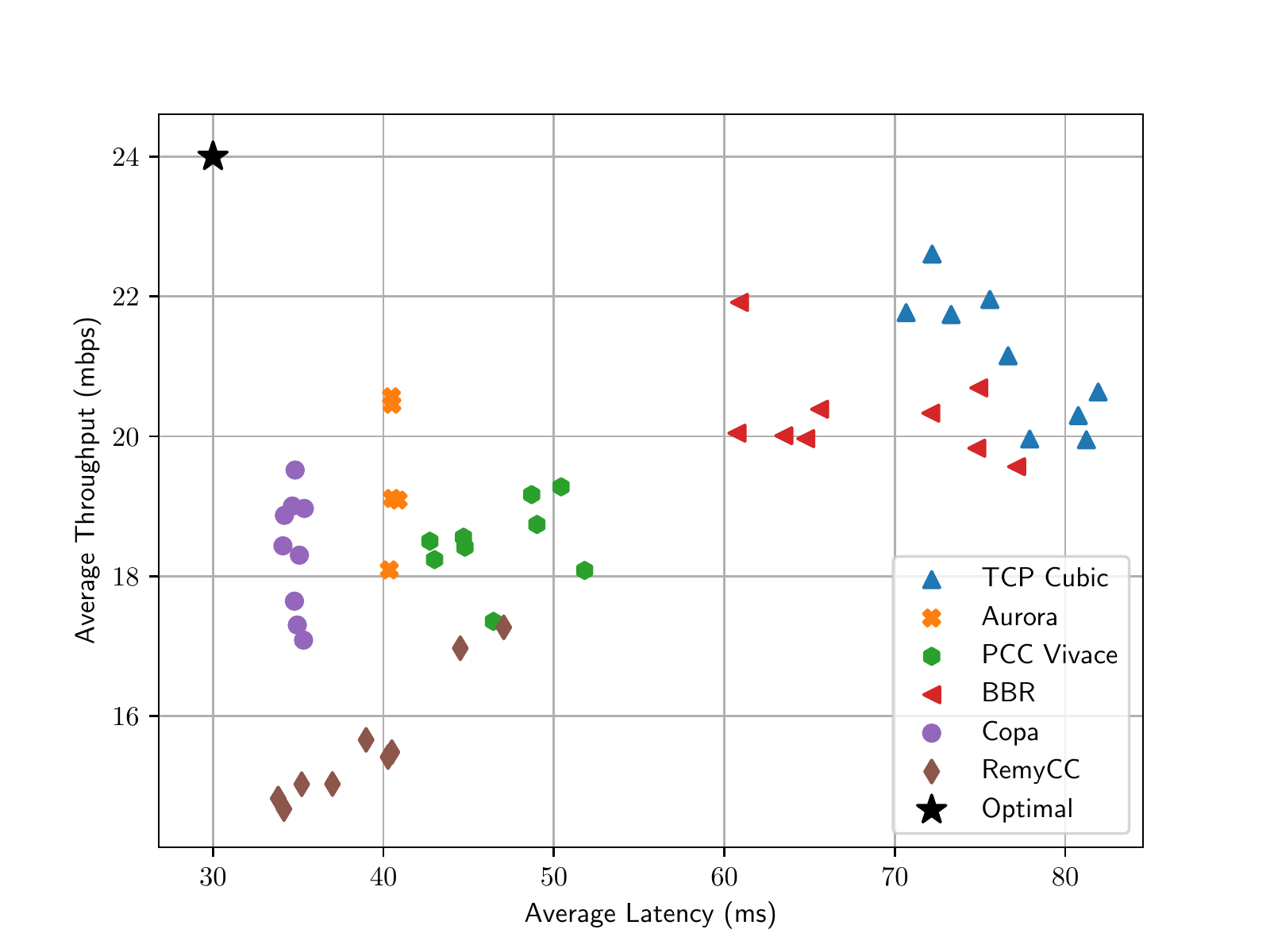}
    \caption{Average throughput and latency on a dynamic link with 32ms of base latency, a 500 packet queue and no random loss.}
    \label{fig:dynamic_link}
\end{figure}

Figure~\ref{fig:dynamic_link} shows as expected, the schemes have different tradeoff points between latency and throughput. TCP CUBIC achieves high throughput at the expense of significantly higher latency and well-documented bad performance in a broad range of network environments (see~\citealt{allegro} and references therein). \name achieves throughput nearly matching BBR~\cite{bbr} but with significantly better latency. Recent designs have worked to achieve lower latency to better support applications like small web queries, real-time video, and gaming. Among the lower-latency schemes, \name achieves both higher throughput and lower latency than PCC-Vivace on average, and higher throughput with similar average latency to RemyCC~\cite{remy}, making it preferable on these dynamic links. Finally, \name comes close to Copa's performance with 4.2\% better throughput and 16\% worse latency.



\subsection{Evaluation Summary}

Our evaluation results demonstrated two key conclusions. First, \name is surprisingly robust to environments outside the scope of its training.  Second, \name is comparable to or outperforms the state-of-the-art in congestion control (BBR, PCC-Vivace, RemyCC, and Copa). This suggests that applying deep RL to congestion control, even with a fairly simple neural network architecture and with fairly limited training, has the potential of outperforming carefully-handcrafted protocols.

\section{A Testbed for RL Congestion Control}
\label{sec:gym_env}

\begin{figure}[!h]
    \centering
    \includegraphics[width=0.9\columnwidth,trim={0 4.5cm 0 8cm},clip]{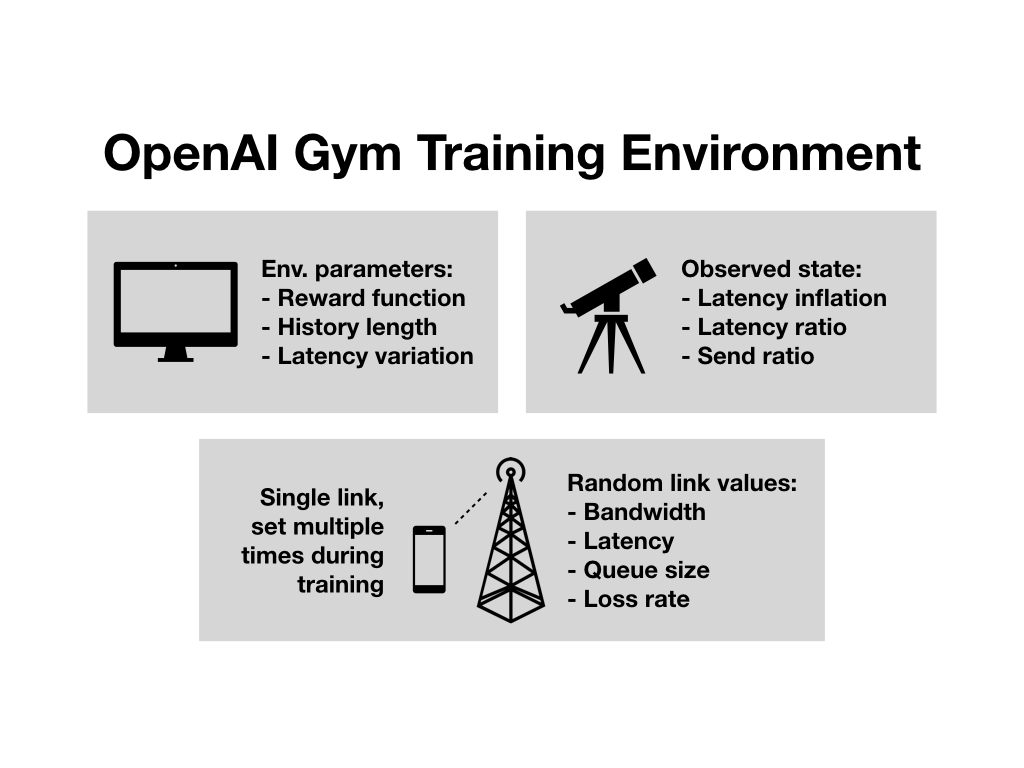}
    \caption{A high-level look at our custom training framework, implemented as an OpenAI Gym environment. By randomizing the link properties during a run, it provides diversified training scenarios. }
    \label{fig:training_diagram}
\end{figure}


Existing network simulators incorporate complex aspects of the network including reachability, routing, packet header parsing, among many other factors that may be useful for other domains, but are irrelevant for congestion control. These extra features complicate and slow down the process of training a congestion control algorithm.
To train \name we implemented a novel simulator, which realistically mimics Internet links with various characteristics. This simulator is sufficiently accurate that models trained in our simulator perform well in the emulated environment used in Section~\ref{sec:evaluation}, which sends real packets using a real Linux stack across virtual network interfaces using standard networking tools.

To open our Internet congestion control environment to the machine learning community in the most standard and accessible way, we provide an OpenAI Gym environment based on our in-house simulator, whose entire code base fits in a few hundred lines of pure Python.
This section covers the basic design and the three abstractions (\emph{links}, \emph{packets} and \emph{senders}) that we use to create a simple simulated network. For a complete description of the environment with source code and documentation, see our \href{https://github.com/PCCproject/PCC-RL}{github repo}.

\begin{table}[h]
    \begin{tabular}{||c c c c||}
        \hline Bandwidth & Latency  & Queue size    & Loss rate \\ [0.5ex] 
        \hline\hline
        100-500 pps & 50-500ms & 2-2981 packets & 0-5\% \\
        \hline
    \end{tabular}
\caption{Training framework variables' ranges, all drawn uniform-randomly, except for the queue size, for which the log is drawn uniform-randomly.}
\label{tab:train_vars}
\end{table}

\subsection{Links}

We simulate links as FIFO queues with the four key parameters identified in Pantheon: \emph{bandwidth}, \emph{latency}, \emph{random loss rate} and \emph{queue size}, each serving the same purpose as in real computer networks. \autoref{tab:train_vars} provides the range of values for these parameters in our example setup.

\subsection{Packets}

Packets normally contain all of the routing information and actual data of network communications, but we don't care about the data, and our routes are vastly simplified. To us, a packet is just a tuple of <sender, latency, is\_dropped>. The sender tells us the route, the latency tells us what the sender will observe when an acknowledgement is returned, and is\_dropped tells us if we should return an acknowledgement.

\subsection{Senders}

Senders are the agents in our networks, each with a \emph{rate}, \emph{route} and list of observations. A sender's rate determines how fast packets are sent on the sender's \emph{route}, a list of links that data and acknowledgements traverse. This route is fixed at network creation time, a reasonably realistic situation, as Internet routes change orders of magnitude more slowly than congestion control decisions are made. Senders then make observations that are given to our machine learning agent. Figure~\ref{fig:training_diagram} gives a list of the observations made by our sender, chosen to include meaningful values used in previous congestion control algorithms~\cite{remy,vivace}.

Using our three objects to create a complete network in our environment is straightforward. The example gym environment in this paper uses just two Links (think two lanes of a street) and single Sender object. The links are initialized with bounded random values for bandwidth, latency, queue size and loss rate given in \autoref{tab:train_vars}. The Sender is given an initial sending rate between 30\% and 150\% of the link bandwidth, and begins taking actions.

\section{Remaining Challenges}

Our results in Section~\ref{sec:evaluation} show that deep RL is indeed capable of learning useful congestion control strategies, which are competitive with dedicated congestion control algorithms. Alongside these promising results, we discuss several aspects of the congestion control problem that are not directly addressed by our RL formulation, and are important for future research. 
We hope that our OpenAI Gym environment would encourage research along these directions.

\subsection{Fairness}

On the Internet, many different congestion control protocols interact, and ours has a significant probability of behaving unfairly in some scenarios (we used neither global optimization as in Remy~\cite{remy}, nor game-theoretic equilibrium analyses as in~\cite{allegro, vivace}). However, our agent might not be friendly towards legacy TCP. Worse yet, if it were trained in an environment where it competed with TCP, it might learn to occasionally cause packet loss just to force TCP into backing off and freeing up network capacity. Can our RL agent be trained to ``play well'' with other protocols (TCP, PCC, BBR, Copa)?




\subsection{Multiple Objectives}

As mentioned earlier, the objective in congestion control is a topic of debate, and users may prefer different tradeoffs between minimizing latency, maximizing throughput, and other performance criteria. Multiobjective RL~\cite{vamplew2011empirical} is a framework for handling the tradeoffs between different reward factors, which in principle could be applied here. We are, however, unaware of any recent deep RL studies in this direction.

\subsection{Generalization and Adaptation to Changes in the Network}

Our results were obtained after training \name on specific network conditions, and we showed that our selection of training network conditions resulted in a congestion control strategy that performed well across a broad range of test conditions. While these empirical results are promising, the question of generalization, or transfer, in RL is an active field of research~\cite{tamar2016value,barreto2017successor,narasimhan2017deep,higgins2017darla}, and providing guarantees about the performance of a congestion control policy when tested outside of its training domain would be valuable for its adoption in practice.

The detection of changes in network conditions is an interesting challenge in itself, and could provide another direction for safe adoption of RL-trained policies. E.g., if we could detect when the network conditions greatly vary from the training conditions, we could fall back to some dedicated congestion control protocol. This problem is an instance of novelty detection~\cite{scholkopf2000support}, and modern approaches employing deep networks could in principle be employed here~\cite{zenati2018efficient,mandelbaum2017distance}.

Another approach is continually adapting the policy to the observed network conditions. Recent approaches in meta-RL~\cite{finn2017model,duan2016rl} could potentially be used to train policies that quickly adapt to network changes.
\section{Related Work}\label{sec:related_work}

\sub{Applications of RL to \emph{highly specialized} contexts.} Several past studies applied RL to highly specialized domains. Researchers proposed RL models for Asynchronous Transfer Mode (ATM) networks~\cite{tarraf1995reinforcement,shaio2005reinforcement}; these predated deep RL and used shallow NNs. In addition, these studies are specially tailored to ATM (a proposed alternative to Internet protocols in general use today); for example,~\cite{tarraf1995reinforcement} requires use of the number of cells in ATM's multiplexer buffer to measure congestion. \cite{hwang2005reinforcement} employs RL to improve congestion control for multimedia networks. This predated deep RL, and also used \emph{cooperating routers in the core network}, where utilization is directly visible, to reach a global equilibrium; in contrast, we target the widely deployed case of transport layer congestion control where senders act \emph{individually with limited information}.


Recently, two studies applied Q-learning~\cite{watkins1992q} to specialized domains: \cite{li2016learning} optimizes TCP for memory-limited IoT devices, and
\cite{silva2016smart} applies Q-learning to delay- and disruption-tolerant networks (DTNs), which apply to scenarios without ongoing end-to-end connectivity such as in interplanetary networks, which does not fit the common case of Internet transport.

\cite{ruffy2018Iroko} applied deep RL to congestion control in \emph{datacenters} and provided an OpenAI Gym environment for benchmarking RL-based congestion control for this context. Unlike Internet congestion control where each sender acts independently, since a datacenter is administered by a single organization, \cite{ruffy2018Iroko} can leverage \emph{global visibility of the network} to optimize a \emph{network-wide reward function}.

\sub{Application of non-deep RL to Internet congestion control.} \cite{CC-ml-kong} presents an RL solver based on the TCP framework. Unlike our work, \cite{CC-ml-kong} employs a rather antiquated RL algorithm (SARSA,~\cite{sutton1998reinforcement}), as opposed to deep RL. In addition, the action space and number of possible reward values are both extremely small (of size $4$), greatly limiting the ability of the model to both learn and react to patterns. Lastly, the RL scheme is compared only to a variant of TCP from 1999 (namely, TCP NewReno~\cite{floyd1999newreno}) and to the Q-learning scheme in~\cite{li2016learning}). 


\sub{Offline optimization of congestion control based on explicit assumptions about the network.} Remy~\cite{remy} is an offline optimization
framework for Internet congestion control. Remy takes as input \emph{explicit} assumptions about
the network, such as specified ranges of link bandwidths, number of competing connections, and more, and also an optimization objective. Then, Remy generates a stochastic model of the network and computes a congestion control protocol that approximates the optimum with respect to this model. Thus, unlike RL, Remy does not learn to react to patterns in traffic and network conditions, but relies on human-specified assumptions. When the prevalent traffic/network conditions deviate from Remy's input assumptions this might result in poor performance.

\sub{Congestion control via online-learning (multi-armed bandits).} PCC~\cite{allegro,vivace} employs online learning techniques to guide congestion control. While this provides valuable \emph{worst-case} guarantees (namely, no regret), unlike \name, it does not learn the prevailing network regularities and adapt rates accordingly. Thus, while PCC does provide robustness, it does not customize to the experienced network conditions, and our evaluation showed scenarios where our approach provides higher performance.


\section{Conclusion and Future Work}\label{sec:conclusion}

We presented \name, a congestion control protocol powered by deep RL. Our evaluation results show that even with fairly limited training, \name is competitive with the state-of-the-art. We highlighted crucial challenges that should be addressed to simplify real-world adoption of such schemes. In future work, we plan to extend our simulation software to include realistic ancillary data and support for multi-agent decision making. 
To facilitate further research on deep-RL-guided congestion control, we open-sourced our code and a test suite based on the OpenAI Gym interface.

\section*{Acknowledgements}

We thank Huawei for ongoing support of our research on congestion control. The fourth author is supported by the Israel Science Foundation (ISF).

\bibliography{main}
\bibliographystyle{icml2019}

\end{document}